\input harvmac

\def\vp{{\bf p}}

\def \ep{\epsilon}

\def \P {{\hat P}}

\def \P {\Phi}
\def \CN {{\cal N}}
\def \II  { II} 

\def \k {\kappa} 
\def \F {{\cal F}}

\def \del {\partial}

\def \chi {\chi}

\def \m {\mu}

\def \vp {\varphi }

\def \o {\omega}

\def \ov {\over }

\def \lr { \lref}
\def\np {{\it Nucl. Phys. }}
\def \pl {{\it  Phys. Lett. }}

\def \prl {{\it  Phys. Rev. Lett. }}

\def \ijmp {{\it Int. J. Mod. Phys. }}


\def\np#1#2#3{Nucl. Phys. {\bf B#1} (#2) #3}
\def\pl#1#2#3{Phys. Lett. {\bf B#1} (#2) #3}
\def\prl#1#2#3{Phys. Rev. Lett. {\bf #1} (#2) #3}
\def\prd#1#2#3{Phys. Rev. {\bf D#1} (#2) #3}

\def\jhep#1#2#3{JHEP {\bf#1}(#2) #3}

\def\ijmp#1#2#3{Int.~J.~Mod.~Phys. {\bf #1} (#2) #3}
\def\atmp#1#2#3{Adv.~Theor.~Math.~Phys. {\bf #1} (#2) #3}



\def\Tr{{\rm Tr}}

\def\vol{{\rm vol}}

\def \C {{\cal C}}


\lref\dgmorb{M.R. Douglas, B.R. Greene and D.R. Morrison,
``Orbifold Resolution by D-branes'', \np{505}{1997}{84},
  {\tt hep-th/9704151}}

\lref\dougegs{M.R. Douglas, ``Enhanced Gauge Symmetry
in M(atrix) theory'', \jhep{007}{1997}{004},
  {\tt hep-th/9612126}}

\lref\jmorb{C.V. Johnson and R.C. Myers, ``Aspects of
Type $\II$B Theory on ALE Spaces'', \prd{55}{1997}{6382},
  {\tt hep-th/9610140}}

\lref\gipol{E.G.~Gimon and J.~Polchinski, ``Consistency Conditions for
Orientifolds and D Manifolds'', \prd{54}{1996}{1667},
  {\tt hep-th/9601038}}

\lref\polten{J.~Polchinski, ``Tensors from K3 Orientifolds'',
\prd{55}{1997}{6423}, {\tt hep-th/9606165}}

\lref\branegeom{H.~Ooguri and C.~Vafa, ``Geometry of $\CN=1$ dualities
in four dimensions'', \np{500}{1997}{62}; {\tt hep-th/9702180}}

\lref\absorb{S. Gubser and I. Klebanov,``Absorption by Branes
and Schwinger Terms in the World Volume Theory,''
\pl{413}{1997}{41}}

\lref\KW{I.R. Klebanov and E. Witten, ``Superconformal Field Theory on
Threebranes at a Calabi-Yau Singularity,'' 
{Nucl. Phys.} {\bf B536} (1998) 199, {\tt hep-th/9807080}.}

\lref\GK{
S.S. Gubser and I.R. Klebanov, ``Baryons and Domain Walls in an
{\cal N}=1 Superconformal Gauge Theory,'' {Phys. Rev.} {\bf D58} 
(1998) 125025, {\tt hep-th/9808075}.}

\lref\MP{D.R. Morrison and M.R. Plesser,
``Non-Spherical Horizons, I,''
Adv. Theor. Math. Phys. {\bf 3} (1999) 1,  {\tt hep-th/9810201}}

\lref\SG{S.S. Gubser, ``Non-conformal examples of AdS/CFT,''
{\tt hep-th/9910117}}

\lref\Gir{L. Girardello, M. Petrini, M. Porrati  and A. Zaffaroni,
``Novel Local CFT and Exact Results on Perturbations of N=4 
Super Yang Mills from AdS Dynamics,''
JHEP {\bf 9812} (1998) 022, 
{\tt hep-th/9810126};
``Confinement and condensates without fine tuning in supergravity
duals  of gauge theories,''
JHEP {\bf 9905} (1999) 026, {\tt hep-th/9903026}
.}

\lref\gppz{L. Girardello, M. Petrini, M. Porrati, A. Zaffaroni,
``The Supergravity Dual of N=1 Super Yang-Mills Theory,''
{\tt hep-th/9909047}}

\lref\bvv{J. de Boer, E. Verlinde and  H. Verlinde,
``On the Holographic Renormalization Group,''
{\tt hep-th/9912012}}

\lref\DZ{J. Distler and F. Zamora,
``Nonsupersymmetric conformal field theories from
stable anti-de Sitter spaces,'' Adv.\ Theor.\ Math.\ Phys.\  {\bf
2} (1999) 1405, {\tt hep-th/9810206};
``Chiral Symmetry Breaking in the AdS/CFT Correspondence,''
{\tt hep-th/9911040}.}

\lref\amp{A.M. Polyakov, ``A Few Projects in String Theory,''
in: \ 
 Les Houches Summer School on Gravitation and Quantizations,
 Session
   57, Les Houches, France, 5 Jul - 1 Aug 1992,  J.
   Zinn-Justin and B. Julia eds. (North-Holland, 1995)
{\tt hep-th/9304146}}

\lr\AP{A.M. Polyakov, ``String Theory and Quark Confinement,''
{ Nucl. Phys. B (Proc. Suppl.)} {\bf 68} (1998) 1, {{\tt
hep-th/9711002}}. }

\lref\Freed{D.Z. Freedman, S.S. Gubser, K. Pilch  and  N.P. Warner,
``Renormalization Group Flows from Holography--Supersymmetry and a c-Theorem,''
{\tt hep-th/9904017}}

\lref\mfour{E.~Witten, ``Solutions of Four-dimensional
  Field Theories via M-theory'', \np{500}{1997}{3}; {\tt hep-th/9703166}}

\lref\aspegs{P.S. Aspinwall, ``Enhanced Gauge Symmetries and K3
Surfaces'', \pl{357}{1995}{329}, {\tt hep-th/9507012}}

\lref\bsvegs{M. Bershadsky, V. Sadov and C. Vafa, ``D-strings on
D-manifolds'',
\np{463}{1996}{398}}

\lref\dixon{L. Dixon, D. Friedan, E. Martinec and S. Shenker,
``The Conformal Field Theory of Orbifolds,''
\np{282}{1987}{13}}

\lref\br{M. Bershadsky and A. Radul,
``Conformal Field Theories with Additional $\IZ_N$ Symmetry,''
\ijmp{A2}{1987}{165}}

\lref\kwdiscrete{L.M. Krauss and F. Wilczek,
  ``Discrete Gauge Symmetries
in Continuum Theories'', \prl{62}{1989}{1221}}

\lref\reidrev{M. Reid, ``McKay correspondence'',
  {\tt alg-geom/9702016}}

\lref\bkv{M.~Bershadsky, Z.~Kakushadze and  C.~Vafa, 
``String expansion as large N expansion of gauge theories",
\np {523} {1998} {59}, {\tt hep-th/9803076}}
\lref\bj{M.~Bershadsky and A.~Johansen, {\tt hep-th/9803249}}

\lr \wol{ O. DeWolfe, D.Z. Freedman, S.S. Gubser  and  A. Karch,
``Modeling the fifth dimension with scalars and gravity,''
hep-th/9909134. }

\lref\KT{I.R.~Klebanov and A.A.~Tseytlin, ``D-Branes and
Dual Gauge Theories in Type 0 Strings,''
\np{546}{1999}{155}, {\tt hep-th/9811035}}

\lref\DM{M.~Douglas and G.~Moore,
``D-branes, quivers, and ALE instantons,'' {\tt hep-th/9603167}}

\lref\jthroat{J.~Maldacena,
``The Large N limit of superconformal field theories and
  supergravity,'' \atmp{2}{1998}{231},
{\tt  hep-th/9711200} }

\lref\gkp{S.S.~Gubser, I.R. Klebanov, and A.M. Polyakov,
  ``Gauge theory correlators from noncritical string theory,''
  \pl{428}{1998}{105}, {\tt hep-th/9802109}}

\lref\EW{E.~Witten, ``Anti-de Sitter space and holography,''
\atmp{2}{1998}{253},
{\tt hep-th/9802150}}

\lref\AMP{A.M.~Polyakov, ``The Wall of the
Cave,''  Int. J. Mod. Phys. {\bf A14} (1999) 645, 
{\tt hep-th/9809057}}

\lref\KS{S.~Kachru and E.~Silverstein, ``4d conformal field theories
and strings on orbifolds,''  \prl{80}{1998}{4855},
{{\tt hep-th/9802183}}.}
\lref\LNV{A.~Lawrence, N.~Nekrasov and C.~Vafa, ``On conformal field
theories in four dimensions,'' \np{533}{1998}{199},
{{\tt hep-th/9803015}}}

\lref\GNS{S.~Gubser, N.~Nekrasov, S.~Shatashvili,
``Generalized Conifolds and 4d N=1 SCFT,'' hep-th/9811230.}

\lref\NS{N. Nekrasov and S. Shatashvili,
``On non-supersymmetric CFT in four dimensions,''
{\tt hep-th/9902110.}, L.~Okun Festschrift, North-Holland,
in press}

\lref\KNS{I.R. Klebanov, N. Nekrasov and S. Shatashvili,
``An Orbifold of Type 0B Strings and Non-supersymmetric Gauge Theories,''
hep-th/9909109}

\lref\JM{J. Minahan, ``Glueball Mass Spectra and Other Issues for
Supergravity Duals of QCD Models,''
JHEP {\bf 9901 } (1999) 020,
 {\tt hep-th/9811156};
 ``Asymptotic Freedom and Confinement from Type 0
String Theory,'' JHEP {\bf 9904} (1999) 007,
{\tt hep-th/9902074}}

\lref\KTnew{I.R. Klebanov and A.A. Tseytlin, ``Asymptotic Freedom and
Infrared Behavior in the Type 0 String Approach to Gauge Theory,''
\np{547}{1999}{143}, {\tt hep-th/9812089} }

\lref\Blum{R. Blumenhagen, A. Font and D. Lust,
``Non-Supersymmetric Gauge Theories from
D-Branes in Type 0 String Theory,''
{\tt hep-th/9906101.}}

\lref\BCR{M. Bill\' o, B. Craps and  F. Roose, ``On D-branes in
Type 0 String Theory,'' {\tt hep-th/9902196.}
}

\lref\Zar{K. Zarembo, ``Coleman-Weinberg
Mechanism and Interaction of D3-branes in Type 0 String Theory, {\tt
hep-th/9901106}}

\lref\ceresole{A. Ceresole, G. Dall'Agata, R. D'Auria, and S. Ferrara,
``Spectrum of Type IIB Supergravity on $AdS_{5}\times {T}^{1,1}$: 
Predictions
On ${\CN}=1$ SCFT's,'' Phys.\ Rev.\  {\bf D61}  (2000)
066001,
{\tt  hep-th/9905226}}

\lref\diac{D.-E. Diaconescu, M. Douglas and J. Gomis, ``Fractional Branes
and Wrapped Branes,'' {JHEP} {\bf 02} (1998) 013,
{\tt hep-th/9712230}}

\lref\Kehag{
A. Kehagias, ``New Type IIB Vacua and Their
 F-Theory Interpretation,''
 Phys.\ Lett.\  {\bf B435} (1998)  337, 
{{\tt hep-th/9805131}}.
}

\lref\SGub{S.S. Gubser, ``Einstein Manifolds and Conformal Field Theories,''
{Phys. Rev.} {\bf D59} (1999) 025006, hep-th/9807164. 
}

\lref\cd{
P.~Candelas and X.~de la Ossa, ``Comments on Conifolds,''
{Nucl. Phys.} {\bf B342} (1990) 246.}

\lref\pope{
D.~N.~Page and C.~N.~Pope,``Which Compactifications Of D = 11 Supergravity Are
Stable?,''
Phys.\ Lett.\  {\bf B144} (1984)  346.}

\lref\Das{
K.~Dasgupta and S.~Mukhi, ``Brane Constructions, Fractional
Branes and Anti-de Sitter Domain Walls,'' 
JHEP {\bf 9907}, 008 (1999), {\tt hep-th/9904131}.}

\lref\repul{C.~Johnson, A.~Peet and J.~Polchinski, 
``Gauge Theory and the Excision of Repulson Singularities,''
{\tt hep-th/9911161}}

\lr\kleb{
S.S. Gubser and I.R. Klebanov, ``Absorption by Branes and Schwinger
Theory,'' {Phys. Lett.} {\bf B413} (1997) 41,
  {{\tt hep-th/9708005}}; for a review see I.R. Klebanov,
``From Threebranes to Large N Gauge Theories,''
{{\tt hep-th/9901018}}. }

\lref \schw{J.~H.~Schwarz,
``Covariant Field Equations Of Chiral N=2 D = 10 Supergravity,''
Nucl.\ Phys.\  {\bf B226} (1983)  269. }
\lref \hull {E.~Bergshoeff, C.~Hull and T.~Ortin,
``Duality in the type II superstring effective action,''
Nucl.\ Phys.\  {\bf B451} (1995)  547, 
{\tt hep-th/9504081}}

\lr \town{K.~Skenderis and P.~K.~Townsend,
Phys.\ Lett.\  {\bf B468} (1999) 46, 
{\tt hep-th/9909070};
K.~Behrndt and M.~Cvetic,
``Supersymmetric domain wall world from D = 5 simple gauged
supergravity,''
{\tt hep-th/9909058} }

\lref \kn{I.R. Klebanov and N. Nekrasov, 
``Gravity Duals of Fractional Branes and Logarithmic RG Flow,''
{\tt hep-th/9911096}.}

\lref \GT{  G.~'t Hooft,
``A Planar Diagram Theory  for Strong Interactions,''
Nucl.\ Phys.\  {\bf B72} (1974)  461.}

\lref\Seif{D.P. Jatkar and S. Randjbar-Daemi,
``Type IIB string theory on $AdS_5 \times  T^{nn'}$,''
Phys. Lett. {\bf B460} (1999) 281, {\tt hep-th/9904187}.}

\lref\cvetic{M. Cvetic, H. Lu and  C.N. Pope,
``Domain Walls and Massive Gauged Supergravity Potentials,''
{\tt hep-th/0001002}.}

\lref\sfet{I. Bakas, A. Brandhuber  and  K. Sfetsos,
``Domain walls of gauged supergravity, M-branes, and algebraic curves,''
{\tt hep-th/9912132}.}

\lref\GHKK{S. Gubser, A. Hashimoto, I. R. Klebanov and  M. Krasnitz,
``Scalar Absorption and the Breaking of the World Volume 
Conformal Invariance,''
Nucl. Phys. {\bf B526} (1998) 393.}

\lref\Angel{C. Angelantonj and  A. Armoni,
``Non-Tachyonic Type 0B Orientifolds, Non-Supersymmetric Gauge Theories and
Cosmological RG Flow,'' {\tt hep-th/9912257}.}

\Title{\vbox
{\baselineskip 10pt
{\hbox{PUPT-1919}\hbox{OHSTPY-HEP-T-00-002}
\hbox{hep-th/0002159}
}}}
{\vbox{\vskip -30 true pt\centerline {Gravity Duals of
Supersymmetric}
\medskip
\centerline {$SU(N)\times SU(N+M)$ Gauge Theories}
\medskip
\vskip4pt }}
\vskip -20 true pt
\centerline{ Igor R.~Klebanov$^{1}$ and Arkady A.
Tseytlin$^{2}$\footnote{$^*$} {Also at Lebedev 
Physics Institute, Moscow and Imperial College, London.}
 }
\smallskip\smallskip
\centerline{$^{1}$ \it Joseph Henry
Laboratories, Princeton University, Princeton, New Jersey 08544}
\centerline{$^{2}$ \it  Department of Physics, 
The Ohio State University,  
Columbus, OH 43210 }
\bigskip\bigskip
\centerline {\bf Abstract}
\baselineskip12pt
\noindent
\medskip
The world volume theory on
$N$ regular and $M$ fractional D3-branes at the conifold
singularity
is a non-conformal ${\cal N}=1$ supersymmetric $SU(N+M)\times
SU(N)$
gauge theory. In previous work the Type IIB supergravity dual of this
theory was constructed to leading non-trivial order in
$M/N$: it is the ${AdS}_5\times { T}^{1,1}$ background
with NS-NS and R-R 2-form fields turned on. 
Far in the UV this dual description
was shown to
reproduce the logarithmic flow of couplings found in the field
theory.
In this paper we study the supersymmetric RG flow at all scales.
 We introduce an ansatz for the 10-d  metric and other fields
 and show that the 
 equations of motion may be derived in first order form from
a simple superpotential.
 This allows us to  explicitly solve for the
 gravity dual of the RG  trajectory.
\bigskip

\Date{02/00}

\noblackbox 
\baselineskip 15pt plus 2pt minus 2pt

\newsec{Introduction}

The AdS/CFT correspondence \refs{\jthroat,\gkp,\EW}
is usually motivated by comparing stacks of
elementary branes with corresponding gravitational backgrounds in string
or M-theory. For example, the correspondence \kleb\ 
between a large number $N$ of coincident D3-branes and the 
3-brane classical solution leads, after
an appropriate low-energy limit is taken, to the duality between
${\cal N}=4$ supersymmetric $SU(N)$ gauge theory and Type IIB strings on
$AdS_5\times S^5$ \refs{\jthroat,\gkp,\EW}. 
This construction gives an explicit
realization of the gauge theory strings \refs{\GT,\AP}.

In order to construct the Type IIB duals
of other 4-dimensional CFT's, one may place the D3-branes at appropriate
conical singularities \refs{\DM,\Kehag,\KW,\MP}. 
Then the background dual to the CFT on the
D3-branes is $AdS_5\times X_5$ where $X_5$ is the Einstein manifold
which is the base of the cone. 
In addition to regular D-branes which can reside
on or off the conical singularity, there are also ``fractional'' D-branes
pinned to the singularity \refs{\gipol,\dougegs}. 
In previous work \kn\
the effect of such fractional branes on the dual supergravity
background was considered, and it was shown how they
break the conformal invariance.
In this paper we continue this line of investigation, and calculate
the back-reaction of the fractional branes on the gravitational
background. We obtain and solve a system of first-order
equations describing renormalization group (RG)
flow in the gravity dual of the ${\cal N}=1$
supersymmetric $SU(N)\times SU(N+M)$
gauge theory. This theory is realized on D-branes at the conifold
singularity and we review it in section 2.

Our study of the gravitational RG flow builds on recent work studying such
flows in gauged 5-d supergravity 
\refs{\Gir,\DZ,\Freed,\wol,\bvv,\cvetic,\sfet}. 
In particular, we will make use of
the results on ${\cal N}=1$ supersymmetric flows 
\refs{\Freed,\gppz,\SG} which reduce 
second-order equations to a much simpler first-order gradient flow
induced by a superpotential function (extensions of these methods to
non-supersymmetric flows were given in \refs{\wol,\town,\bvv,\amp}).
In our example we start with an ansatz for the 10-d background and reduce
it to a 5-d gauged supergravity coupled to scalar fields. This gives
a clear geometrical meaning to the scalars fields, so that we can
follow the RG evolution of the entire 10-d background.
This is similar in spirit, although not in detail, to examples
of RG flow found in 10-d  type 0 string theory
 \refs{\KT,\JM,\KTnew,\Angel}.
An interesting novel 
feature  of the solution we find is that 
its existence crucially 
depends on the presence of the Chern-Simons 
term in the type IIB supergravity action.

\newsec{ RG Flow Associated with 
Fractional Branes on the Conifold}

The conifold is a singular Calabi-Yau manifold
described in terms of complex variables
$w_1,\dots, w_4$ by the equation \cd\ 
 $
 \sum_{a=1}^4 w_a^2 = 0.
$
The base of this cone is 
${ T}^{1,1}= (SU(2)\times SU(2))/U(1)$
whose Einstein metric may be written down explicitly 
as follows \refs{\pope,\cd},
\eqn \co {
 d s_{T^{1,1}}^2=
{1\over 9} \bigg(d\psi + 
\sum_{i=1}^2 \cos \theta_i d\phi_i\bigg)^2+
{1\over 6} \sum_{i=1}^2 \left(
d\theta_i^2 + {\rm sin}^2\theta_i d\phi_i^2
 \right)
\ .}
The ${\cal N}=1$
superconformal field 
theory on $N$ regular D3-branes placed at the singularity
of the conifold has gauge group $SU(N)\times SU(N)$ and global
symmetry $SU(2)\times SU(2)\times U(1)$  \KW\
which is a  symmetry of the metric  \co. The theory contains
two chiral superfields
$A_i$  transforming as $(N,\overline N)$ 
and as a doublet of the first
$SU(2)$, and 
two  chiral superfields
$B_k$ transforming  as $(\overline N, N)$ and as
 a doublet of the second
$SU(2)$. The R-charge of all four chiral superfields is $1/2$ and the
theory has an exactly marginal superpotential
${\cal W} =\epsilon^{ij} \epsilon^{kl}\Tr A_iB_kA_jB_l$.

Type IIB supergravity modes on $AdS_{5}\times{ T}^{1,1}$ have been matched in
some detail with operators in this gauge theory whose dimensions
are of order $1$ in the large $N$ limit \refs{\SGub,\Seif,\ceresole}. 
Topologically, $T^{1,1}$ is $S^2\times S^3$ so that additional objects
may be constructed from wrapped branes \GK.
In particular, a D5-brane wrapped over the 2-cycle
acts as 
a domain wall in $AdS_{5}$. 
If this domain wall is located
at $r=r_w$ then, by studying the behavior of wrapped D3-branes upon
crossing it, it was shown in \GK\ that for $r> r_w$ the gauge group
changes to $SU(N+1)\times SU(N)$.
This is precisely the gauge theory expected on $N$
regular and one fractional D3-branes.
 Thus, a D5-brane wrapped over the
2-cycle is nothing but a fractional D3-brane placed at a definite
$r$.
The identification of a fractional D3-brane with a wrapped D5-brane
is consistent with the results of \refs{\jmorb,\dougegs,\diac,\Das}.

Adding $M$ fractional D3-branes thus produces $SU(N+M)\times SU(N)$
supersymmetric gauge theory
coupled to the chiral superfields $A_i$ and $B_k$. 
Its supergravity dual carries $M$ units of 
the R-R 3-form, $H_{RR}$, flux through
the 3-cycle of $T^{1,1}$. In \kn\ it was shown that this flux
induces a radial variation of the
integral of the NS-NS 2-form potential 
 $\int_{S^2} B_{2}$.
This was interpreted as the stringy dual of the logarithmic RG flow
in the field theory. In the next section we set up the gravitational RG
flow equations systematically, so that the back-reaction of the
3-form field strengths on other fields may be calculated. This will allow
us to follow the flow far into the infrared and address the issues
related to singularities. 


\newsec{ The Supergravity Ansatz and the Effective Action}

The type IIB supergravity equations \schw\  can be 
obtained from the action
\eqn\efc{  S_{ 10} 
= - {1\ov 2\k_{10}^2}  \int d^{10} x \bigg( \sqrt{-g_{10}} \big[ \ R_{10} 
 - { \textstyle{1\ov 2}} (\del \P)^2
- { \textstyle{1\ov 12}} e^{-\P}   (\del B_2)^2 
} 
$$ -  \  { \textstyle{1\ov 2}}  e^{2 \P} (\del \C)^2 
 - { \textstyle{1\ov 12}} e^{ \P}  (\del C_2  - \C  \del B_2) ^2
- { \textstyle{1 \ov 4\cdot 5!}}  F^2_5\ \big]
- { \textstyle{1\ov 2\cdot 4! \cdot (3!)^2    }}
 {\ep_{10}} C_4 \del C_2 \del B_2 + ... \bigg) \ ,  $$
$$(\del B_2)_{MNK} \equiv 3 \del_{[M} B_{NK]}\ , \ \
\  \  \ (\del C_4)_{MNKLP} \equiv 5 \del_{[M} C_{NKLP]} \ , $$ 
$$ F_5= \del C_4 + {5} (B_2 \del C_2 - C_2 \del B_2) \  , $$
with the additional on-shell constraint $F_5 = *F_5$
\hull.
In \kn\ these equations were solved to leading order in $M/N$, 
and it was
shown that the back-reaction on the metric and 5-form 
fields 
enters at
order $(M/N)^2$. To study the back-reaction, we introduce
the following ansatz.
The 10-d Einstein frame metric will be chosen as 
a sum of the  5-d space-time metric  and 
the internal 5-manifold metric which 
has the same symmetries  as \co:
\eqn\mett{
 ds^2_{10} = L^2 \bigg[  e^{- { 2 \ov 3} 
 (B+ 4 C)} ds^2_5 + ds^2_{5'} \bigg]\ ,  }
\eqn\coop{ ds^2_{5'} = 
{1\over 9} e^{2 B} \bigg(d\psi + 
\sum_{i=1}^2 \cos \theta_i d\phi_i\bigg)^2+
{1\over 6}e^{2C} \sum_{i=1}^2 \left(
d\theta_i^2 + {\rm sin}^2\theta_i d\phi_i^2 \right)
\ .}
Here $B,C$ are in general functions of the 
5-d space-time coordinates 
and the conformal factor of the 5-d 
metric is chosen to preserve the Einstein frame upon compactification
to 5-d. The numerical coefficients $1/9$ and $1/6$
(which are the same as in \co) 
are inserted in order to have $B=C=0$ for the
$AdS_5\times T^{1,1}$ solution.
$L$ is the scale related to the radius  of 
$AdS_5$ ($L^4 \sim N$).

We set the {{RR}} scalar $\C$ to zero 
(this will be consistent with
the ansatz for 3-form fields made
below)
and study
the case where
  \eqn\add{
ds^2_5  = du^2 + e^{2 A(u)} dx_n dx_n \ , }
and $B,C$ and the 10-d dilaton $\P$ are functions of $u$.
Following \kn\ we shall note that since 
the fractional D3-brane, i.e. the wrapped D5-brane, creates
R-R 3-form flux through ${T}^{1,1}$, \  $H_{\rm RR}
= dC_2$ should be
proportional to the closed 3-form. This 3-form was constructed in \GK,
\eqn\har{
H_{{\rm RR}}= \  P\ e^\psi \wedge \o_2 \ , \ \ \ \ \ 
\o_2 \equiv { 1 \ov \sqrt 2} ( e^{\theta_1} \wedge e^{\phi_1} - 
e^{\theta_2} \wedge e^{\phi_2})
 \ .
}
Here $P$ is a constant proportional to the integer number 
$M$ of $H_{RR}$ flux units.
In the  normalization 
where  $L=1$ which we shall use below, $P \sim M/N$
($N$ is fixed by the boundary conditions at $u=u_0$).
We have introduced the orthonormal basis of
1-forms \GK
\eqn\basis{ e^\psi =
{1\over 3} (d\psi + \sum_{i=1}^2 \cos \theta_i d\phi_i )\ ,
\qquad e^{\theta_i} = {1\over \sqrt 6} d\theta_i\ ,
\qquad e^{\phi_i} = {1\over \sqrt 6} \sin \theta_i d\phi_i\ .
}
The form of  the NS-NS 2-form potential is  also as in \kn
\eqn\ansa{
B_2  = \  T(u) \ \o_2 \ ,\ \ \ \ \ \  
\ \ \  H_{\rm NS} = \ T'(u) \   du \wedge \o_2
\ . }  
$T$ plays the role of 
a scalar field in the effective 5-d supergravity
theory.

The natural ansatz for the self-dual 5-form is 
$$
F_5 = \F + *\F \ , $$ 
\eqn\sela{ \F= K(u) \vol ({\rm T}^{1,1})\ =  \ K(u) \ 
e^\psi\wedge e^{\theta_1}\wedge e^{\phi_1}\wedge e^{\theta_2}
\wedge e^{\phi_2}\ ,  }
 $$*\F = e^{4A- { 8\ov 3} ( B + 4 C) }\ K \  du \wedge 
 dx^1 \wedge  dx^2 \wedge  dx^3 \wedge  dx^4\ .$$
Then  $H_{\rm RR}$  in \har\  satisfies 
the required equation $d*H_{\rm RR} \sim F_5\wedge H_{\rm NS}$ since 
$F_5\wedge H_{\rm NS}=0$.

Note that our ansatz
preserves the symmetry between the two 2-sphere 
factors in \coop. Had we put in different warp
factors for the two 2-spheres, there would still
be a solution where they are equal. This is due to the
special form of the ansatz for the 3-form field strengths.
Thus, our ansatz is rigidly constrained and, as we will see, leads to
rather simple RG equations.

The $F_5$ equation of motion  
implies a relation between 
the functions $K$ and $T$.
Indeed, $d*F_5 = dF_5\sim  H_{\rm NS} \wedge H_{\rm RR}$ implies
that the scalars $K$ and $T$ are not independent:  
\eqn\impl{
K'= P T'\ , \ \ \ {\rm i.e.} \  \ \  \  K(u) = Q + P T(u) \ . }
For $P=0$ the constant $Q$ plays the role 
of the 5-brane charge, $Q \sim N$.
For non-zero $P $ the constant $Q$ can be absorbed into a
redefinition of the function  $T$.
This follows from the fact that 
$F_5= d C_4 + 5(B_2dC_2 - C_2 dB_2)
= dC'_4 + 10 B_2dC_2 $, where $C_4'$ is related by a field redefinition to
$C_4$, 
$$ C_4= C_4' + 5 B_2 C_2\ .
$$
Since $d(dC_4')=0$, $dC_4'$ must contain the 
volume 5-form 
$\vol ({\rm T}^{1,1})$ part with {\it constant} coefficient $Q$.

The equation for $T$ which follows
from $ d *H_{\rm NS} \sim F_5\wedge H_{{{\rm RR}}}$ 
has the structure
 $ \nabla^\mu (e^{-\Phi}\nabla_\mu T ) \sim  P (Q + P T) $.

In general, the  dilaton will be running according to
\eqn\dill{
 \nabla^2 \Phi ={1\over 12} (e^\Phi H_{{\rm RR}}^2 - e^{-\Phi} H_{\rm NS}^2)
\ .}
However, as we shall explain below, 
there exists a special class of solutions for which $\P$ 
remains constant. Then
$H_{\rm NS}^2 = e^{2 \Phi} H_{{{\rm RR}}}^2$, or 
 $ T' = P e^{ \Phi }e^{ - { 4 \ov 3} (B+C)}$,
i.e. $T$ can be expressed in terms of 
  $B$ and $C$.

The full  set of  equations for 
the  5-d metric function $A$ and  scalars $B,C,\P,T$
can be found from the following 5-d action 
which can be obtained 
 from  \efc\ by taking into account 
the solution of the $F_5$ equation of
motion
  and integrating by
parts 
\eqn\efcp{  S_{ 5} 
= - {2\ov \k_{5}^2}  \int d^{5} x 
\ \sqrt{g_{5}} \bigg[ 
{ 1 \ov 4}  R_5  
- { 1 \ov 2} G_{ab}(\vp)  \del \vp^a \del \vp^b 
- V(\vp)\bigg]
 \ . }
Here the scalar fields $\vp^a=(q,f,\P,T)$ include 
the  ``diagonal" combinations of  $B$ and $C$
\eqn\dia{q = { 2 \ov 15} (B + 4C) \ , 
\ \ \ \  \ f= -{ 1 \ov 5} (B- C) \ , }
 which measure 
  the volume
  and the ratio of scales of the internal manifold
  \coop. In the
presence of a 5-d cosmological constant
both  turn out to be positive (mass)$^2$ scalars (see below).
Explicitly, in  \efcp\  
\eqn\cti{
G_{ab}(\vp)  \del \vp^a \del \vp^b
= 15 (\del q)^2  + 10 (\del f)^2 
 + {1\ov 4} (\del \P)^2
  + { 1 \ov 4}  e^{-\Phi- 4f - 6 q}(\partial T)^2 \ , 
} 
\eqn\pott{
V(\vp) = e^{ - 8 q}  ( e^{ -12 f } - 6 e^{- 2 f}) 
+ {1\over 8} P^2 e^{\Phi + 4 f - 14 q}
 + {1\over 8} (Q +  P T)^2 e^{ - 20 q} 
\ .}
The kinetic term for $T$ comes from the $e^{-\P}   (\del B_2)^2 $
term in \efc.
The three terms in the potential  have the following origin.
The first combination in \pott\ comes from 
$R_{10}$ in \efc\ 
and reflects the curvature of the internal space \coop\
present in the limit of  constant ``radii" $e^B$ and $e^C$.
The second term is the $e^{ \P}  (\del C_2)^2$ 
evaluated on the solution \har.
The third term corresponds to  $F^2_5$.
Note that the contribution of the Chern-Simons
term in \efc\ is already effectively  accounted for
 (it should not  directly 
influence the 10-d gravitational part of  equations of motion).
As discussed above, 
and as apparent  from the structure of the action 
\cti,\pott,  
for  non-zero  $P$ the constant $Q$  can be absorbed 
into $T$.
This is an important feature of the system under
 consideration.
  
When written 
in terms of the 5-d metric function
$A(u)$  and the scalars $\vp^a(u)$
the action takes the following  form 
\eqn\fcrp{  S_{ 5} 
= - {2{\rm Vol}_4 \ov \k_{5}^2}  \int du
\ e^{4 A} \bigg[ 3 A'^2 
- { 1 \ov 2} G_{ab}(\vp)  \vp'^a  \vp'^b -
 V(\vp)\bigg]
 \ . }
The set of equations obtained by varying $A,q,f,\P,T$
should be supplemented  by the ``zero-energy'' constraint
\eqn\zeren{
3 A'^2 
- { 1 \ov 2} G_{ab}(\vp)  \vp'^a \vp'^b + 
 V(\vp) = 0 \ .}
The simplest ``fixed-point" solution found for $P=0$ 
corresponds to the $AdS_5$ space:
when all scalars are constant, 
$V= -5 + {1\over 8} Q^2$.
In the normalization where $L=1$ this  gives the  
$AdS_5$ space of unit radius, i.e.   $A=u$,   
for  $Q = 4$. 

The reader may be slightly puzzled by the origin of the 
rescaling that sends $N\rightarrow Q$ and $M\rightarrow P$ 
(recall that $M$ is the actual number of $H_{RR}$ quanta).
If we reinstate the dependence on $L$ and
  $g_s=e^{\Phi}$ then 
 the kinetic term for $T$ in \cti\ 
 has a factor of $e^{-\P}/L^4$, 
 while the $N^2$ and $(N+MT)^2$ terms in the potential
 have factors $ e^{\P}/L^4$  and $1/L^{8}$
  respectively.  Noting that the scale of the  string-frame
  10-d metric, which is to be held fixed,
   is related to the scale of the Einstein-frame
  metric, $L\sim N^{1/4}$,  by $L_s =(g_s)^{1/4} L 
   \sim  (g_s N)^{1/4}$, 
    we find that these three factors
   become $1/L_s^4$, 
   $  L_s^4/N^2$ and  $1/N^2$  respectively.
   That explains why $M$ becomes replaced by $P\sim M/N$ while
$N$ by $Q\sim 1$.

\newsec{ The First Order System and its Solution}

In general, the action $S_5$ \fcrp\ generates a  
system of second-order differential equations.
However, as  was observed in \refs{\Freed,\gppz}, 
in the case of 
solutions preserving some amount of supersymmetry
this system  can be replaced  by 
 {\it first}-order equations in $u$:
\eqn\firr{
\vp'^a = { 1 \ov 2} G^{ab} { \del  W \ov \del \vp^b} \ ,
\ \ \ \ \ \ \ \ 
A' = - { 1 \ov 3} W (\vp) \ , }
where the superpotential $W$ is a 
function of $k$ scalars $\vp^a$
satisfying
\eqn\sati{
V = {1\over 8} G^{ab} {\partial W\over \partial \vp^a}
{\partial W\over \partial \vp^b} - {1\over 3} W^2
\ . 
}
It is easy to check directly that \firr,\sati\ imply the
second-order equations following from 
\fcrp,\zeren.
This first-order form leads to a crucial simplification in finding the
RG flow.\foot{
It was noted in \refs{\wol,\town,\bvv} that,
even in the absence of supersymmetry, 
any solution of the second-order 
system of equations that follows from \fcrp\
can be obtained  as  solution of the 
{ first}-order system \firr,\sati.
Given $V(\vp)$, one may, in principle, solve
the non-linear equation \sati\ for $W$.
The solution is not unique, depending on $k$
integration constants. In general, in the absence of supersymmetry,
$W$ is not guaranteed to be simple (in particular, monotonic).}

Since the RG trajectory we are after is dual to ${\cal N}=1$
supersymmetric gauge theory, we expect the equations of motion to
be simply expressible in the first order form.
We have succeeded in guessing a simple superpotential $W$ which governs
these equations
(for definiteness, we assume $Q,P \geq 0$):
\eqn\spec{
W= -e^{ - 4 q}  ( 2e^{ - 6 f } + 3  e^{4 f}) 
 +  {1\over 2} (Q +  P T) e^{ - 10 q} \ .}
For $Q=4,\ P=0$ we get  for small $f$ and $q$
\eqn\www{  W= -3  - 60 f^2 + 60 q^2  + ... \ , }
which shows that $f$ and $q$ are a ``good"  (diagonal) 
choice of fields.
The potential $V$ \pott,\sati\ has the expansion
\eqn\vvv{V=-3 + 60 f^2   + 240 q^2 + ...  \ , }
which shows that these fields have masses 
$m_q^2=32$ and $m_f^2=12$. 
The  combination 
$q$ in \dia\ is the standard fixed scalar
degree of freedom \GHKK\ corresponding to the overall volume of the
compact manifold. For the KK reduction on $S^5$, the
superpotential for the fixed scalar was derived in \cvetic.

Although we have not checked explicitly that the first-order flow 
generated by the superpotential \spec\  preserves
${\cal N}=1$ supersymmetry, we believe that it is the case.
One indirect check of the supersymmetry is to set $P=0$
and to consider the linearized equation for $q$,
$$ q'= 4 q\ .$$
Its solution is $q\sim e^{4u}$ which corresponds to adding a source
for an operator of dimension 8 \refs{\gkp,\EW}
(schematically, this operator has
a $\Tr\ F^4$ structure \GHKK). 
This describes the leading perturbation from
the $AdS_5\times T^{1,1}$ background toward 
the metric of $N$ D3-branes at the
conifold, which is known to preserve ${\cal N}=1$ supersymmetry.
In section 6 we exhibit the full BPS 3-brane solution, which serves
as a consistency check on the first-order equations we have derived.

Note that, although $V$ in \pott\ depends on 
the dilaton $\P$, 
the superpotential $W$ does not depend on it.\foot{
The $\P$ dependence of $V$ is reproduced 
in \sati\ due to the dilaton 
dependence of the ``metric" $G_{ab}$ entering the kinetic term of $T$
in \cti.}
As a result, the dilaton remains constant along the flow!
In what follows we shall set $\P=0$.

The system of equations for $T,f,q$ that follow from 
\firr,\spec\ is thus:
\eqn\syse{
T' =    P  e^{- 4 q  + 4 f  } \ , }
\eqn\yse{
 f' = - { 3 \ov 5}  e^{- 4 q + 4 f } ( 1- e^{ - 10 f })  \ , }
\eqn\ssse{
q' =   { 2 \ov 15}    e^{- 4 q + 4 f } ( 3 + 2 e^{ - 10 f }) 
 -  { 1 \ov 6}  (Q +  P T) e^{ - 10 q}
 \ . }
Note that for $P=0$ this system has a stable fixed point 
solution $f=0,\ q={ 1 \ov 6} \ln (Q/4), \ T={\rm const}$
mentioned above. It
corresponds to the $AdS_5\times T^{1,1}$ background with 
unit radius, and $q=0$ if $Q=4$. Therefore, we will use $Q=4$ below. 
 
To fix the boundary conditions, we assume that above a UV
cut-off scale $u=u_0$ we have the
 superconformal theory with $P\sim
M=0$.
Physically this corresponds to inserting 
$M$ fractional anti-branes
at $u=u_0$. Thus we set
 \eqn\inii{ q=f=0\ ,\qquad T=T_0 \ \ \ \ \ {\rm at}\ \  \ \  u=u_0\ , }
 and consider the solution for $u < u_0$. 
The constant $T_0$ determines the value of $g_1^{-2} - g_2^{-2}$ at the
UV cut-off \refs{\KW,\kn}
($g_1,g_2$ are the two gauge coupling constants).
Since  $\del W/\del f$ (the r.h.s. of  \yse)
 vanishes for $f=0$, we find  that $f(u)=0$
along the entire RG flow.
 This means that the shape of the internal
manifold $T^{1,1}$ does not change, only its overall size does!

This simplifies our task to finding just two
functions, $T(u)$ and $q(u)$. The first order equations for them are governed
by the superpotential
\eqn\newsuper{
W= - 5 e^{ - 4 q}   
 +  {1\over 2} (Q +  P T) e^{ - 10 q} \ ,}
which is \spec\ with $f$ set to zero. Thus, we have
\eqn\sysenew{
T' =    P  e^{- 4 q } \ , }
\eqn\sssenew{
q' =   { 2 \ov 3}    e^{- 4 q }  
 -  { 1 \ov 6}  (Q +  P T) e^{ - 10 q}
 \ . }
Introducing the variables
\eqn\newvar{ K(u) = Q +  PT(u)\ , \qquad Y(u) = e^{6q(u)}
\ , 
}
we get from  \sysenew\ 
\eqn\xder{
K'=  P^2 e^{-4 q} =  P^2 Y^{-2/3}\ .
}
Using also \sssenew\
we find that
\eqn\crucial{ {d Y \over d K} = {1\over P^2} (4 Y - K)
\ .
}
This has a general solution
\eqn\rgsol {Y =\  a_0 \ e^{4 K/P^2} + {K\over 4}  + {P^2\over 16}
\ .
}
The constant $a_0$
has to be chosen to implement the UV boundary
condition that $Y=1$ when $K=K_0= 4+ PT_0$: 
\eqn\inti{ a_0= -\left ({P^2\over 16}+ {PT_0\over 4}\right ) 
\exp \left [-{16\over P^2}- {4 T_0\over P} \right ] \ .
}
This completely fixes the relation between $T$ and $q$ along the
RG trajectory. In particular, for small $T-T_0$,\ 
$q= - { 1 \ov 6} T_0 (T-T_0)  + \ldots\ .$
This is consistent with the perturbative solution
of \syse,\ssse\ 
\eqn\eee{
T = T_0+  P (u-u_0)  + \ldots \ , \ \ \ \ \ \
q = -{1\over 6} P T_0  (u-u_0) + \ldots
\ . }
Note that $u-u_0$ translates into $\ln(\Lambda/\Lambda_0)$ 
in the field theory. Thus, the variation of $T$ translates into 
a logarithmic flow of $g_1^{-1} - g_2^{-2}$
in the field theory confirming the result of \kn. 
Furthermore, we can now calculate higher order corrections to the
metric and $T$ in powers of $P\sim M/N$.

Substituting \rgsol\ into \xder\ we find
\eqn\rgeq{
K' = P^2 \left [ {P^2\over 16}
+ a_0 e^{4 K/P^2}  +{K\over 4} \right ]^{-2/3}
\ ,
}
which in turn leads to an implicit equation for $K(u)$,
\eqn\impl{
u_0 - u= {1\over P^2} \int^{K_0}_K dz
\left [ {P^2\over 16}
+ a_0 e^{4 z/P^2} +{z\over 4} \right ]^{2/3}
\ .
}
Using this relation and \rgsol\ we also have a
 relation between $q$
and $u$. 

To complete our solution, we need to find $A(u)$.
The equation for the function $A$ in \firr\
has the form 
\eqn\rrr{
A' =  { 1 \ov 3} \bigg[
e^{ - 4 q + 4 f }  ( 3 + 2e^{ - 10 f } ) 
 -  {1\over 2} (Q +  P T) e^{ - 10 q} \bigg]
=   q' + { 1\ov P } T'  + {2 \ov 3} f' \ , }
where we have used \syse--\ssse\
to express the exponents in terms of the  derivatives. Thus 
 {\it in general}  $A$ is   simply a linear combination
\eqn\kli{
A(u) = A_0 + q(u)   + {2 \ov 3} f(u) + { 1\ov P } T(u)\ .
}
For our particular trajectory with $f=0$, this gives $A$
in terms of $q$ and $T$. 
The integration constant $A_0$ may be shifted by a rescaling 
of  4-d coordinates $x_n$ in \add.
We can choose it so that the metric \add\
approaches the canonical $AdS_5$ one, i.e. 
$A(u) \rightarrow u$ for $u \to u_0$.
The resulting expression for the 10-d metric \mett,\coop,\add\
 may be written as ($L=1$)
\eqn\newmett{ ds^2_{10}  = 
 e^{-5 q} du^2 + e^{2A- 5q} dx_n dx_n + e^{3q}
ds_{T^{1,1}}^2\ ,
}
where we have used that for $f=0$  eq. \dia\ gives 
   $B=C= {3\ov 2} q$.
Introducing the coordinate $y$ such that 
$dy = e^{-(A-A_0)} du$, we may write this metric in the form
\eqn\meew{
ds^2_{10} = e^{-3 q  + {2\ov P} T} (dy^2 + dx_n dx_n)  + e^{3 q}
ds^2_{T^{1,1}} \ . }

\newsec{Solution in a Special Case}

There is a simple choice of the boundary
 condition for $T$ in \inii, $T_0 = -P/4$,
which leads to $a_0=0$. Then an explicit solution
 of the RG equations
is straightforward: \rgsol\ becomes  
\eqn\sicn{ Y= {K\over 4} + {P^2\over 16}
\ ,
} 
and we find from \xder\ that
\eqn\newrg{Y' = {P^2\over 4} Y^{-2/3} \ , 
}
i.e. 
\eqn\newsol{  Y(u) = e^{6q(u)} = 
a_1  P^{6/5} (u-u_s)^{3/5} \ ,\ \ \ \ \  a_1 = (5/12)^{3/5}\ .
}
Then we have
\eqn\kbeh{ K= 4+ PT(u)  = - {1\ov 4} P^2  + 
    4 a_1 P^{6/5} (u-u_s)^{3/5} 
 \ . 
}
Obviously, $u_s$ is the position of the 
singularity where $T^{1,1}$
shrinks to vanishing size. 
To find the  relation of $u_s$ to the UV cut-off $u_0$
we note that the boundary condition 
 \inii\  implies $Y(u_0)=1$, i.e.  
\eqn\reeq{ u_0 - u_s = {12\over 5 P^2 }\ .
}
The effective scale factor in 5-d gauged supergravity 
metric \add\ is given by (see \kli) 
\eqn\kik{ e^{2A}\  \sim\  P^{2/5} (u-u_s)^{1/5} 
\exp [ 8 a_1   P^{-4/5} (u-u_s)^{3/5}]
\ .
} 
The fact that this vanishes at $u=u_s$ seems to indicate the presence of
a naked singularity in the geometry. However, recall
that in the 10-d metric \newmett\ 
the effective  scale factor in front of $dx_n dx_n$ is
\eqn\yty{ e^{2A- 5q} \ \sim\  P^{-3/5} (u-u_s)^{-3/10} 
 \exp [  8 a_1 P^{-4/5} (u-u_s)^{3/5}]
\ ,
}
which,  in fact,  blows up at $u_s$.
Thus, in order to study the singularity
structure, it is essential to know the full 
10-d  form of the solution.

\newsec{More General Solutions}

In this section we go beyond the `near-horizon' approximation and
construct asymptotically flat solutions of the first-order system of 
equations \firr. For $P=0$ our solution describes regular D3-branes
at the conifold singularity. For $P\neq 0$ we find an interesting
generalization of this solution with a logarithmically running charge.

Let us look for 10-d metric of  the following 4+6 form
\eqn\mew{
ds^2_{10} =   s^{-1/2}(r)   dx_n dx_n 
 +  h^{1/2} (r)  (dr^2 + r^2 ds^2_{T^{1,1}} ) \ . }
The relation of this to the notation of \newmett\ is
\eqn\rewq{ s^{-1/2}(r) = e^{2 A(u)- 5 q (u)}\ ,
\ \ \ \ \ \  \   e^{3 q(u)/2 }  = r  h^{1/4} (r)\ .  } 
The new radial coordinate, $r$, is related to $u$ through
\eqn\coordrel{
e^{-4 q(u)} du = { dr\ov r } \ . 
}
Hence, multiplying the last two equations, we have
\eqn\vimp{
h^{1/4} (r) dr = e^{-5 q(u)/2}  du \ ,
}
which shows that \mew\ is equivalent to \newmett.
 
Using \sysenew\ we get
\eqn\gett{
dT  =    P  e^{- 4 q } du =  P d (\ln r) \ , }
which has a solution 
\eqn\rett{
T(r) = \tilde T + P \ln r \ .  } 
We already know the solution \rgsol\ for $q(T)$, hence we find
\eqn\finn{
r^4 h  (r) = 
e^{6 q }
 =     a_0 \ e^{4 \tilde Q/P^2 + 4 \ln r } 
    + {1 \over 4} (\tilde Q + P^2 \ln r )
  + {1\over 16} P^2 
\ , }
where  $$\tilde Q \equiv  Q + P \tilde T\ . $$
Thus, we have the following explicit solution
for $h(r)$:
\eqn\nhf{
h(r) =  b_0 +  { k_0 + P^2 \ln r  \over 4r^4}
\ ,}
where
$$
k_0= \tilde Q + { 1 \ov 4} P^2 \ , \ \ \ \ \ \ \ 
b_0 = a_0 \ e^{4 \tilde Q/P^2} \ . 
$$
To solve for $s(r)$ we note that according to
\rewq,\kli
\eqn\uio{ s (r) = e^{-4T(r)/P + 6 q (r)}
    =   { 1 \over r^4} e^{6 q(r)} =  h  (r)  \ . }
Remarkably, the 10-d metric thus  assumes the 
usual ``D-brane" form
\eqn\New{
ds^2_{10} =   h^{-1/2}(r)   dx_n dx_n 
 +  h^{1/2} (r)  (dr^2 + r^2 ds^2_{T^{1,1}} )\ .  }
The function
 $h(r)$, given in \nhf, may be written as
\eqn\mayb{
h(r)= b_0  +  { P^2 \ln {r\over r_*}  \over 4 r^4}\ ,
}
where $r_*$ is defined by 
\eqn\uyo{
- P^2 \ln  r_* =  Q + P \tilde T + { 1 \ov 4} P^2
\ .}
The Ricci scalar for this 10-d metric
turns out to have  has the following
simple form
\eqn\rscal{
R= -  { 5 r^{-1}  h'(r) + h''(r) \ov 2 h^{3/2} (r) }=
 { 4 P^2     
\over  \left ( 4b_0 r^4 + {P^2} \ln{r\ov r_*} \right )^{3/2}}
\ . }
This makes it clear that the metric becomes singular at $r=r_s$
such that $h(r_s)=0$.

For  $P=0$ and $b_0>0$
the  solution \New\ is precisely the BPS metric
 of D3-branes placed at the
conifold singularity (note that \rscal\
 vanishes in this case, as it should). 
This shows that,
at least in this case, the solution to 
the first order system preserves
supersymmetry.

 For $P\neq 0$, the solution is still
asymptotically flat. Note in particular that
the NS-NS field strength falls off at large $r$, 
$ H_{NS}= {P\over r} dr\wedge \omega_2$. A remarkable property of 
the solution \New\ is that it
looks like a threebrane metric whose 
effective charge and mass per unit volume
depend on the radius logarithmically. 
To strengthen this interpretation, let us consider a test D3-brane
oriented along the source D3-branes and placed at radial coordinate $r$.  
 As in the  case where 
  the transverse 6-d CY space in \New\
 is replaced by $R^6$, the gravitational force
 on the test brane is proportional to the derivative 
  of the metric function  \nhf,  
  \eqn\tens{
  \m (r) =   - r^5 { d h\ov dr} =  
   K(r) \ , \ \ \ \ \  
   K(r) = Q + P T(r)= \tilde
  Q + P \ln r \ . }
 For this reason $K(r)$ may be thought of as the mass per unit
 volume enclosed inside radius $r$.
 Note that this is different from the 
  coefficient of the $1\ov 4r^4$ term in 
 \nhf\  (i.e. $ K(r) + {1 \ov 4}  P^2$), 
but is the same as the coefficient in the 5-form field strength \sela. 
  This is in  agreement with the
   expected BPS nature of  this configuration.
Indeed, the force on the static D3-brane probe
oriented parallel to the source brane  will 
vanish as a result of the balance between the electric 
force proportional to the 5-form component $\cal F$ in \sela\
and the gravitational force proportional to \tens.
The cancellation of forces is also another argument
in favor of our solution preserving ${\cal N}=1$ supersymmetry.

To study the solution in more detail, 
we have to distinguish the cases where
$b_0$ is positive, zero, or negative. The asymptotically flat region
exists only if $b_0>0$. In this case, and for $P=0$, we find the
AdS horizon at $r=0$. For $P\neq 0$, however, the situation
 is completely
different because the enclosed 3-brane  charge or mass density
at radius $r$ is $K(r)$.  
As $r$ decreases, so does this  effective 3-brane charge density.
 At $r=r_e > r_*$   where $K(r_e)=0$, the  gravitational force changes
 sign, i.e. inside this
radius we have antigravity.
Thus, $r=r_e$ is similar to the enhancon radius found
 in a different
setting in \repul.
 It is not hard to check that the metric is
nonsingular at $r=r_e$ (this is  obvious from \rscal).
 Continuing to $r< r_e$ past $r_*$  we eventually reach
the singularity where $h(r_s)=0$, 
i.e. $4b_0 r^4_s = {P^2} \ln{r_*\ov r_s}$.
 One may speculate that this singularity
should be ``excised'' because it occurs in the region where the
effective 3-brane charge and tension are negative. 

The case of $b_0=a_0=0$ is the one discussed in 
the preceding section.
The exact metric found there is simply \New\ expressed in terms of a 
different radial coordinate.
Again, in  this case the singularity 
occurs at $r$ smaller than $r_e$, the point where 
the effective 3-brane charge per unit volume vanishes.
Indeed, from  \kbeh\ we see that
$ K = - { 1 \ov 4} P^2$  is negative  at the singularity. 
 
Finally,  let us consider the case $b_0 < 0$. Now $r$ cannot increase
indefinitely: as $r$ increases we find 
a singularity at $r_+$ where $h$
vanishes. For very small $r$ there is another
curvature singularity located at $r_- > r_*$.
 Thus, for $b_0<0$
there are two curvature singularities, and we have
$ r_* < r_- < r_+$.
It is not clear, however, if the case $b_0<0$ 
is physical: if we interpret
the metric \New\ as the geometry around $N$
 regular and $M$ fractional
D3-branes placed at the conifold singularity,
then this geometry
is required to have the asymptotically flat region at large $r$.

\newsec{ Gauge Theory Interpretation and Comments}

In this section we summarize some main points and 
further discuss the 
gravitational
 solution  dual to the RG flow in  the
supersymmetric $SU(N+M)\times SU(N)$
gauge theory.
In investigating the actual solution, we first consider the
case $ P\sim M/N \ll 1$. Then 
we see from \eee\ that near the UV
cut-off $u_0$ both 
derivatives $T'$ and $q'$ are small, hence the gravity
approximation is valid.
As $u$ decreases from $u_0$, $K=Q+ PT$
decreases monotonically.
At the value $u_e$ given by
\eqn\uyu{   u_e = u_0 - 
{1\over P^2} \int^{K_0}_0 dz
\left [ {P^2\over 16}
+ a_0 \exp (4 z/P^2) +{z\over 4} \right ]^{2/3}
} 
$K$ reaches $0$. Since the 5-form field
 strength vanishes at $ u_e $, this location is similar to 
the enhancon radius found in \repul.
As already mentioned above, for
 $u < u_e$ we find `antigravity,' and it is plausible
to assume that this region has to be excised in a full string theoretic
treatment.

If we continue the effective gravity solution to $u < u_e$, we find a
 singularity of the metric:
the value $u_s$ where $Y= e^{6q}=0$. Since 
$e^{-16/P^2}$ is negligible for small $P$, we can see from
\rgsol\ that $ K(u_s) \approx - P^2/4$.
Using \syse\ and \ssse\ we can derive the behavior of the metric
functions $A$, $q$ 
and $K$. The leading behavior coincides with that found
in the exact solution exhibited in section 5:
near  the singularity both $q$ and $A$   diverge to $-\infty$
as ${1\over 10}\ln (u-u_s)$. 
While  $e^{2A}$ vanishes,  we note again that 
 in the 10-dimensional
metric \newmett\ the conformal factor is not $A$, but rather 
\eqn\rath{
A-{5\over 2}q \approx -{3\over 20} \ln (u-u_s)
\ ,
}
i.e. 
  the longitudinal part of the 10-d
metric expands rather than contracts
 as we approach 
$u_s$. 
The 10-d metric \newmett\ is nevertheless
singular because of the volume of $T^{1,1}$ shrinking to
zero. In particular, the 10-d Ricci scalar is  
$ R_{10} \sim 
[P^2 (u-u_s)]^{-3/10}$.
This is why the gravity approximation near $u_s$ has
 to be taken with
a grain of salt: stringy corrections could alter the conclusions
entirely. We find it suggestive, however, that far in the infrared
the compact 5-manifold seems to be removed dynamically -- this is
a desirable feature for understanding the dynamics of realistic gauge theories
\AMP. Perhaps one day it will be possible to understand the effective
5-d string theory where the singular compact manifold is `integrated
out.' 

The fate of the singularity is an interesting issue. From
 \rgsol\ we see that, for all $a_0> -P^2/16$, the singularity
is hidden behind the enhancon-type locus $K(u)=0$ 
where the effective 3-brane
tension vanishes. Thus, following \repul\ we may conjecture that
such singularities have to be excised in a string-theoretic treatment.
Since negative $a_0$ may be unphysical, this protection of
singularities may be a general phenomenon in the system we are
considering.

Another issue we need to address is the fact that a change
of $T$ shifts the effective 3-brane charge.  Given that 
 $T$ is scale
dependent, it therefore appears that far in the IR the gauge group is
different from that found in the UV.
Let us suggest the following qualitative 
picture.  Since $Q$ scales as $N$, and $P$ scales as $M$, 
from the point of view of 
the dual $SU(N+M)\times
SU(N)$ gauge theory, we conjecture that $N$ starts decreasing dynamically
as the theory flows to the IR.
At first,  the variation of $T$ may be interpreted as 
the variation of 
$g_1^{-2} - g_2^{-2}$ in the field theory \kn.
But what happens when
we reach a value of  $u$ where one of the couplings diverges?
Since a shift of $T$ corresponds to a shift of $Q$, 
the natural continuation past this infinite coupling involves the
field theory with $N$ replaced by $N-M$. Repeating this reasoning many
times we seem to eventually arrive at a theory with $N$ comparable to
$M$ or even at  the theory with a single gauge group $SU(M)$.
Presumably, this is the theory described by the vicinity of $ u_e$
where $K$ is near 0. This is an intriguing conjecture, but of course
we need further checks to establish it, even on a qualitative level,
because of difficulties with the effective gravity approximation.

In view of the above, it appears that, even if we start
with $M\ll N$, the theory dynamically drives itself
into a regime where $N$ and $M$ are comparable. A natural question then
is: why can't we start with $P\sim M/N$ of order one from the beginning.
Then the problem is that, even in the UV,  the flow is no longer slow
and the supergravity approximation is suspect. 
Nevertheless, it might
provide a useful qualitative picture.
Even if $P$ is of order 1, the solution typically exhibits a repulson
singularity hidden behind the enhancon-type locus, similar to those found in
\repul.

\bigskip
\noindent
{\bf Acknowledgements}
\bigskip
We are grateful to R. Corrado, N. Nekrasov, K. Pilch and N. Warner for
useful discussions.
The work of I.~R.~K. was supported in part by the NSF
grant PHY-9802484 and by the James S. McDonnell
Foundation Grant No. 91-48.
The work of A.T. is supported in part by
the DOE grant  DOE/ER/01545, PPARC SPG grant   PPA/G/S/1998/00613, 
 EC TMR grant  ERBFMRX-CT96-0045, 
INTAS grant No.96-538,
and NATO grant PST.CLG 974965.

\vfill\eject
\listrefs
\end